\documentclass[letterpaper,twocolumn,10pt]{article}
\usepackage{usenix2020_SOUPS}

\usepackage{tikz}
\usepackage{amsmath}

\begin{document}

\date{}

\title{\Large \bf Surveying Vulnerable Populations:\\
  A Case Study of Civil Society Organizations}

\def\plainauthor{Nikita Samarin, Alisa Frik, Sean Brooks, Coye Cheshire, Serge Egelman}

\author{
{\rm Nikita Samarin}\\
UC Berkeley / CLTC
\and
{\rm Alisa Frik}\\
ICSI / UC Berkeley
\and
{\rm Sean Brooks}\\
UC Berkeley / CLTC
\and
{\rm Coye Cheshire}\\
UC Berkeley
\and
{\rm Serge Egelman}\\
ICSI / UC Berkeley
} 

\maketitle
\thecopyright

\begin{abstract}
Compared to organizations in other sectors, civil society organizations (CSOs) are particularly vulnerable to security and privacy threats, as they lack adequate resources and expertise to defend themselves. At the same time, their security needs and practices have not gained much attention among researchers, and existing solutions designed for the average users do not consider the contexts in which CSO employees operate. As part of our preliminary work, we conducted an anonymous online survey with 102 CSO employees to collect information about their perceived risks of different security and privacy threats, and their self-reported mitigation strategies. The design of our preliminary survey accounted for the unique requirements of our target population by establishing trust with respondents, using anonymity-preserving incentive strategies, and distributing the survey with the help of a trusted intermediary. However, by carefully examining our methods and the feedback received from respondents, we uncovered several issues with our methodology, including the length of the survey, the framing of the questions, and the design of the recruitment email. We hope that the discussion presented in this paper will inform and assist researchers and practitioners working on understanding and improving the security and privacy of CSOs.
\end{abstract}

\section{Introduction}

Researchers and practitioners have traditionally focused on the security needs and practices of the average users, sidestepping underrepresented and vulnerable communities. In recent years, there has been an increase in research examining the security and privacy behaviors of such groups, revealing nuanced and community-specific concerns and practices that differ from those of the average users~\cite{boyd2014s, mcgregor2016individual, yarosh2013shifting, matthews2017stories, blackwell2016lgbt, guberek2018keeping}. One such vulnerable online population consists of employees working for civil society organizations (CSOs), which include a wide range of groups, such as humanitarian organizations, labor unions, advocacy groups, indigenous peoples movements, faith-based organizations, community groups, professional associations, foundations, think tanks, charitable organizations, and other non-governmental and not-for-profit organizations~\cite{worldbank2020}. Compared to other sectors, civil society operates in elevated-risk contexts, as they are often targeted for political or ideological reasons by state-sponsored actors~\cite{lipton2016perfect}, political opponents~\cite{scott2017bitter}, hate groups~\cite{brandom2016anonymous}, and radicalized individuals~\cite{glaser2020bail}. Whereas attacks against average users and for-profit organizations mostly result in financial losses~\cite{bissell2019ninth}, attacks against individuals working for politically-vulnerable CSOs often carry greater ramifications, including, in severe cases, threats to freedom of expression, liberty, and even life~\cite{marczak2014governments, marczak2015hacking, crete2014communities, deibert2009tracking}.

Prior research indicates that civil society groups lack the funds and human resources to defend themselves against security and privacy threats~\cite{brooks2018defending}. For instance, they maintain a low ratio of IT staff to non-technical staff~\cite{hulshof2017tenth}, do not conduct vulnerability assessments~\cite{cohnreznick2017governance}, and do not adopt solutions aimed at improving their cybersecurity~\cite{sierra2013digital}. Findings from a 2018 report by the Public Interest Registry~\cite{public2018global} indicate that CSOs rarely have access to purpose-built systems and instead, tend to use commodity tools that are not tailored to their needs and elevated risk profiles. For instance, 58\% of surveyed CSOs use Facebook messenger, which is not encrypted by default, to communicate sensitive information~\cite{public2018global}. Although recent attempts have been made to design security solutions that are tailored specifically to CSOs~\cite{googleapp, lerner2017confidante}, they often fail to capture the needs, practices, and mental models of their intended users~\cite{mcgregor2015investigating, marczak2017social}.

In November and December of 2019, we conducted an anonymous online survey with 102 CSO employees to collect information about their risk perceptions and self-reported mitigation strategies. We surveyed employees at a broad range of organizations based in the US, some of which could be classified as high-risk and others as low-risk, in order to compare their average perceived risks. The survey also measured several other factors that we believe affect the attitudes and intentions of individuals to engage in protective behavior (e.g. risk awareness, self-efficacy, perceived support, etc). We also collected organizational and demographic information. Finally, we asked the participants to provide us feedback at the end of the survey to help us improve future iterations of the study.

In the rest of this article, we discuss the motivation for our work, the methodology for the preliminary study, the challenges we faced, and the lessons we learned from our experience and the respondents' feedback. We hope that this discussion will also benefit other researchers and practitioners working in this area. 

\section{Motivation}

One could try to dismiss cyberattacks against civil society and elevated-risk users as ``edge cases'' that deserve less attention than more sophisticated technical attacks, or threats that affect broader user populations. However, there are several reasons why understanding the context in which CSOs operate is essential. First, employees working for CSOs constitute a sizable proportion of the population. In the US alone, they account for 11.4 million jobs or 10.3\% of the non-public sector workforce~\cite{bureau2014nonprofits}. Second, as most CSOs employ standard tools used by millions of users~\cite{public2018global}, while their online risks are amplified compared to the general population~\cite{brooks2018defending}, this particularly vulnerable population could be considered ``extreme users''~\cite{djajadiningrat2000interaction}. Therefore, understanding how CSO employees use mainstream tools could reveal insights on usability and security issues that might be overlooked in the studies with typical user communities. Such insights will help to improve the design of technology for the average use-case as well. For instance, enabling key security choices by default in popular platforms would improve the security outcomes both for the high-risk and the average users. Finally, cyberattacks that target CSOs today are precursors of threats that could affect broader user groups tomorrow~\cite{scott2016security}. Understanding how to protect against such threats for the high-risk users would, therefore, also confer security for the average users. 

\section{Methodology}

\begin{figure*}
\begin{center}
\includegraphics[width=14.5cm]{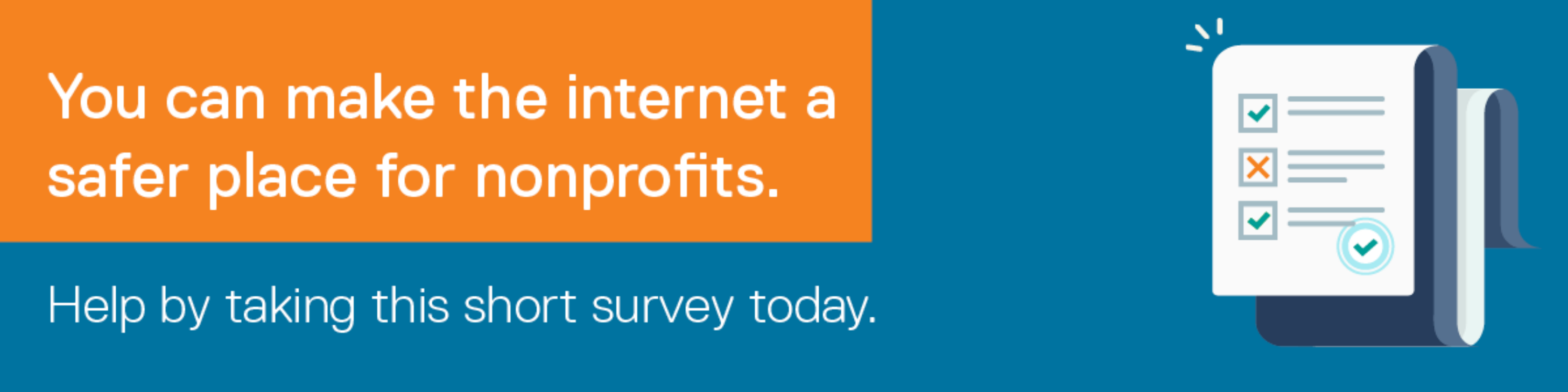}
\end{center}
\caption{Recruitment banner included in the email newsletter.}
\label{fig:banner}
\end{figure*}

The goal of our study is to better understand cybersecurity concerns and practices in CSOs to improve their resilience against cybersecurity attacks. Based on our personal experience working with CSOs and prior work with journalists~\cite{mcgregor2016individual}, activists~\cite{marczak2017social}, and humanitarian workers~\cite{le2018enforcing}, we identified the following seven threats: phishing, malware, online harassment, online reputation attack, physical device compromise, surveillance, and attacks on online services. We designed and executed a survey among employees at CSOs to assess the respondents' risk perceptions of each of these seven threats and to collect information on their self-reported risk mitigation strategies for one specific threat chosen at random. Additionally, the survey presented a list of strategies that correspond to best practices in mitigating each threat and asked participants to report their level of familiarity with each and whether they have made use of them.\footnote{These mitigation strategies were compiled from various online resources that provide security and privacy guidelines for individuals and organizations, which were then reviewed by our research team.}

In addition to following standard questionnaire design guidelines, such as designing scales to measure the constructs, minimizing survey response time, and protecting the confidentiality of responses, we addressed several considerations that were specific to employees at CSOs: establishing trust, preserving anonymity, and recruitment challenges.

\textbf{Establishing trust.} While communicating research risks is essential in any study that involves human subjects, it is especially important when surveying vulnerable populations. More specifically, revealing information about the current security practices and priorities of a CSO could place their employees and the organization as a whole at heightened risk.  While our survey was completely anonymous and did not collect identifiers of any kind, we also had to ensure that our respondents felt safe enough to provide information related to our research goals. In addition to communicating our commitment to anonymization in the consent form, we separately highlighted the anonymity of the survey in a separate location in the survey itself, in order to establish trust with respondents and to relieve their concerns. This also increased the chance that respondents were aware of the safeguards employed, even if some of them did not read the consent form entirely. 

\textbf{Using anonymity-preserving incentive strategies.} We wanted to provide some incentives to respondents as a compensation for their time, and to increase participation rates in the survey. As we are not assigning any identifiers to the survey participants, we cannot follow-up with them to provide any direct compensation. Instead, at the end of the survey respondents can select one of three charities, to which we will proportionally donate our compensation budget at the end of the study. Research has shown that material incentives (either monetary or non-monetary)~\cite{goritz2006incentives} and sharing result summaries~\cite{dillman2011mail} increase response and completion rates for online surveys, but without affecting the quality of responses~\cite{gritz2004impact, sanchez2010analysis}. 

\textbf{Using a trusted intermediary for participant recruitment.} In order to reach our target population, we partnered with TechSoup, a nonprofit that coordinates an international network of other nonprofits, providing technical support, training, and tools.\footnote{\url{https://www.techsoup.org/}} We distributed our survey in one of TechSoup's periodic newsletters, which allowed us to leverage their large reach among nonprofits and their existing connection to our target audience. The survey was promoted via a banner ad (see Figure~\ref{fig:banner}), which included an anonymous link to our survey. The content and format of the banner had to accommodate the existing conventions established in TechSoup's newsletters. Recruitment text with a stronger call-to-action may have attracted more attention, but this compromise was well worth it to directly access our target population. Additionally, we are able to disseminate key findings back through TechSoup's platform so that participants who participated anonymously can review and learn from the results of the study.

\section{Challenges}

Previous surveys run by TechSoup among their network achieved from 2,500 to 20,000 responses, and we were optimistic that we would see at least several thousand participants respond to our survey. Unfortunately, only 160 individuals navigated to the survey in the first place; 16 potential participants were excluded because they never progressed past the consent form, 39 participants were excluded because they did not finish the survey, and 3 respondents were additionally excluded due to incorrect responses to attention check questions, leaving 102 complete responses that we analyzed. After analyzing the valid responses and the feedback from respondents, we identified several issues that we intend to address in the next iteration of the survey, including the survey length, incorrect terminology, non-applicable questions, and the design of the banner ad.

\textbf{Length of the survey.} By far the most common feedback that we have received was that the survey was too long and/or contained repetitive questions (mentioned by 41\% of respondents who provided feedback). To assess key factors that affect the attitudes and intentions of participants to engage in protective behavior, we included a total of 11 scales developed to measure constructs such as risk awareness, self-efficacy, response efficacy, perceived support, perceived culture, and others, each ranging from 2 to 7 items each. This also resulted in a 19.8 minutes median completion time, which is longer than the recommended 10 minutes for online surveys~\cite{revilla2017ideal}.

To address this challenge, we suggest narrowing down the scope of research questions to reduce the number of constructs measured by the survey, or to use the between-subject approach, i.e. present only a subset of questions to each participant if the expected sample size is large. We also suggest providing feedback to respondents as they are completing the survey, for instance, by displaying their current progress in the survey and explaining that although questions might seem repetitious, they are in fact measuring different factors.

\textbf{Incorrect terminology.} Another issue that we discovered was the usage of the word `employee' throughout the survey to refer to the participant. In the beginning of the survey, we clarify that we make no distinction between different capacities of involvement with the organization, and use the word `employee' only for brevity. Nevertheless, 12\% of respondents who provided feedback mentioned that they felt confused when responding to questions as their organization has no or few employees, and is composed mostly of volunteers. 

To avoid this issue, it is important to remember the fact that individuals engage with CSOs in different capacities, including employees, contractors, and volunteers. When referring to the respondent directly, one approach to solving this problem is to use neutral phrasing to encompass anyone working at a CSO (e.g. ``as someone working for a civil society organization, consider the following…''). For questions that involve the organization itself, we recommend exhaustively listing different options to avoid any confusion (e.g. ``How many individuals (including employees, volunteers, contractors, etc.) currently work for your organization?''). 

\textbf{Non-applicable questions.}  Our aim was to survey a broad range of CSOs, regardless of the cause they support, their position in the industry, or their size, which meant that some questions had to be broad enough to cover all types of organizations. For this reason, however, 17\% of respondents who provided feedback mentioned that they were not able to answer some of the questions as they were not applicable to them. For instance, some respondents mentioned that their organization was too small to have information security policies, or that they do not report to the IT department because they are the IT department. For our single- and multiple-choice questions, we only included a `Don't Know' option and, thus, failed to account for questions that do not apply to certain respondents. 

To address this problem, we recommend including a `Not Applicable' option for each question, alongside an optional free-text box that can be used to provide additional comments and clarifications. 

\textbf{Recruitment banner design.} We also identified an issue with the recruitment banner included in TechSoup's email newsletter, as shown in Figure 1. Some members of our research team who had access to the body of the email reported that it was not clear whether the banner had to be clicked to navigate to the survey, and others reported it resembling an advertisement banner that would be ignored. We did not have much control over the design recruitment banner, as it followed TechSoup's design guidelines, but we believe it might have contributed to non-response bias, especially if it disproportionately affected those who did not understand that the banner was clickable or those who assumed that the banner was simply an advertisement and ignored it.

To increase the number of prospective respondents, we recommend highlighting the fact that the banner is clickable, either by including a textual explanation (e.g. ``Click here to participate''), or a graphic that looks like a clickable button on the banner. We also suggest including supporting text in the body of the email newsletter that brings attention to the survey, emphasizes that the survey is for academic and not commercial reasons, sets expectations about the required time commitment, and highlights any compensation offered for participation.

\section{Conclusion}

We applied survey-based methods to understand cybersecurity concerns and practices of CSO employees, including their perceived risks of different security and privacy threats, and their self-reported mitigation strategies. The design of our preliminary survey accounted for the unique requirements of our target population by establishing trust with respondents, using anonymity-preserving incentive strategies, and distributing the survey with the help of TechSoup. However, by carefully examining our methods and the feedback received from respondents, we uncovered several issues with our methodology, including the length of the survey, our usage of terminology, non-applicable questions, and the design of the recruitment banner. We hope that the discussion of these challenges will not only assist us in the design of our future studies but will also benefit other researchers and practitioners working on understanding and improving the security and privacy of CSOs.

\section*{Acknowledgments}

We would like to thank the Citizen Clinic at the Center for Longer-Term Cybersecurity (CLTC) and members of the Berkeley Laboratory for Usable and Experimental Security (BLUES) lab for their support and for providing expert input and review of our survey instruments. We also thank TechSoup for their collaboration and providing access to their network of nonprofits during this research project. This research is sponsored by funding from the CLTC at UC Berkeley.

\bibliographystyle{plain}
\bibliography{references}

\begin{thebibliography}{10}

\bibitem{cohnreznick2017governance}
Cohnreznick 2017 not-for-profit governance and financial management survey.
\newblock Technical report, CohnReznick, 2017.
\newblock Retrieved June 25, 2020 from
  \url{https://www.cohnreznick.com/insights/2017-not-for-profit-governance-financial-management-survey}.

\bibitem{public2018global}
2018 global {NGO} technology report.
\newblock Technical report, Public Interest Registry and Nonprofit Tech for
  Good, 2018.
\newblock Retrieved June 25, 2020 from
  \url{https://www.givingtuesday.org/lab/2018/03/2018-global-ngo-technology-report}.

\bibitem{bissell2019ninth}
Kelly Bissell, Ryan LaSalle, and Paolo~Dal Cin.
\newblock Ninth annual cost of cybercrime study.
\newblock Technical report, Accenture and Ponemon Institute, 2019.
\newblock Retrieved June 25, 2020 from
  \url{https://www.accenture.com/us-en/insights/security/cost-cybercrime-study}.

\bibitem{blackwell2016lgbt}
Lindsay Blackwell, Jean Hardy, Tawfiq Ammari, Tiffany Veinot, Cliff Lampe, and
  Sarita Schoenebeck.
\newblock {LGBT} parents and social media: Advocacy, privacy, and disclosure
  during shifting social movements.
\newblock In {\em Proceedings of the 2016 CHI conference on human factors in
  computing systems}, pages 610--622, 2016.

\bibitem{boyd2014s}
Danah Boyd.
\newblock {\em It's complicated: The social lives of networked teens}.
\newblock Yale University Press, 2014.

\bibitem{brandom2016anonymous}
Russell Brandom.
\newblock Anonymous groups attacked {Black Lives Matter} website for six
  months.
\newblock {\em The Verge}, 2016.
\newblock Retrieved June 25, 2020 from
  \url{https://www.theverge.com/2016/12/14/13951762/anonymous-black-lives-matter-ddos-attack-six-months-hacktivism}.

\bibitem{brooks2018defending}
Sean Brooks.
\newblock Defending politically vulnerable organizations online.
\newblock Technical report, Center for Long-Term Cybersecurity (CLTC), 2018.
\newblock Retrieved June 25, 2020 from
  \url{https://cltc.berkeley.edu/defendingpvos/}.

\bibitem{bureau2014nonprofits}
U.S. Department of~Labor Bureau~of Labor~Statistics.
\newblock Nonprofits account for 11.4 million jobs, 10.3 percent of all private
  sector employment.
\newblock {\em The Economics Daily}, 2014.
\newblock Retrieved June 25, 2020 from
  \url{https://www.bls.gov/opub/ted/2014/ted_20141021.htm}.

\bibitem{crete2014communities}
Masashi Crete-Nishihata, Jakub Dalek, Ronald Deibert, Seth Hardy, Katharine
  Kleemola, Sarah McKune, Irene Poetranto, John Scott-Railton, Adam Senft,
  Byron Sonne, and Greg Wiseman.
\newblock Communities @ risk: Targeted digital threats against civil society.
\newblock Technical report, Citizen Lab, Munk Centre for International Studies
  and University of Toronto, 2014.
\newblock Retrieved June 25, 2020 from \url{https://targetedthreats.net/}.

\bibitem{deibert2009tracking}
Ronald~J Deibert, Rafal Rohozinski, A~Manchanda, Nart Villeneuve, and GMF
  Walton.
\newblock Tracking ghostnet: Investigating a cyber espionage network.
\newblock Technical report, Citizen Lab, Munk Centre for International Studies
  and University of Toronto, 2009.
\newblock Retrieved June 25, 2020 from
  \url{https://citizenlab.ca/2009/03/tracking-ghostnet-investigating-a-cyber-espionage-network/}.

\bibitem{dillman2011mail}
Don~A Dillman.
\newblock {\em Mail and Internet surveys: The tailored design method--2007
  Update with new Internet, visual, and mixed-mode guide}.
\newblock John Wiley \& Sons, 2011.

\bibitem{djajadiningrat2000interaction}
John~Partomo Djajadiningrat, William~W Gaver, and JW~Fres.
\newblock Interaction relabelling and extreme characters: methods for exploring
  aesthetic interactions.
\newblock In {\em Proceedings of the 3rd conference on Designing interactive
  systems: processes, practices, methods, and techniques}, pages 66--71, 2000.

\bibitem{glaser2020bail}
April Glaser.
\newblock Bail organizations, thrust into the national spotlight, are targeted
  by online trolls.
\newblock {\em NBC News}, 2020.
\newblock Retrieved June 25, 2020 from
  \url{https://www.nbcnews.com/tech/tech-news/bail-organizations-thrust-national-spotlight-are-targeted-online-trolls-n1226321}.

\bibitem{googleapp}
Google.
\newblock Advanced protection program, 2019.
\newblock Retrieved June 25, 2020 from
  \url{https://landing.google.com/advancedprotection}.

\bibitem{goritz2006incentives}
Anja~S G{\"o}ritz.
\newblock Incentives in web studies: Methodological issues and a review.
\newblock {\em International Journal of Internet Science}, 1(1):58--70, 2006.

\bibitem{gritz2004impact}
Anja~S Gritz.
\newblock The impact of material incentives on response quantity, response
  quality, sample composition, survey outcome and cost in online access panels.
\newblock {\em International Journal of Market Research}, 46(3):327--345, 2004.

\bibitem{worldbank2020}
World~Bank Group.
\newblock Civil society policy forum, 2020.
\newblock Retrieved June 25, 2020 from
  \url{https://www.worldbank.org/en/events/2020/04/17/civil-society-policy-forum}.

\bibitem{guberek2018keeping}
Tamy Guberek, Allison McDonald, Sylvia Simioni, Abraham~H Mhaidli, Kentaro
  Toyama, and Florian Schaub.
\newblock Keeping a low profile? {T}echnology, risk and privacy among
  undocumented immigrants.
\newblock In {\em Proceedings of the 2018 CHI Conference on Human Factors in
  Computing Systems}, pages 1--15, 2018.

\bibitem{hulshof2017tenth}
Robert Hulshof-Schmidt.
\newblock The tenth annual nonprofit technology staffing and investments
  report.
\newblock Technical report, NTEN, 2017.
\newblock Retrieved June 25, 2020 from
  \url{https://www.nten.org/article/your-guide-to-nonprofit-it-investment/}.

\bibitem{le2018enforcing}
Stevens Le~Blond, Alejandro Cuevas, Juan~Ram{\'o}n Troncoso-Pastoriza, Philipp
  Jovanovic, Bryan Ford, and Jean-Pierre Hubaux.
\newblock On enforcing the digital immunity of a large humanitarian
  organization.
\newblock In {\em 2018 IEEE Symposium on Security and Privacy (SP)}, pages
  424--440. IEEE, 2018.

\bibitem{lerner2017confidante}
Ada Lerner, Eric Zeng, and Franziska Roesner.
\newblock Confidante: Usable encrypted email: A case study with lawyers and
  journalists.
\newblock In {\em 2017 IEEE European Symposium on Security and Privacy
  (EuroS\&P)}, pages 385--400. IEEE, 2017.

\bibitem{lipton2016perfect}
Eric Lipton, David~E Sanger, and Scott Shane.
\newblock The perfect weapon: How {R}ussian cyberpower invaded the {US}.
\newblock {\em The New York Times}, 13, 2016.
\newblock Retrieved June 25, 2020 from
  \url{https://www.nytimes.com/2016/12/13/us/politics/russia-hack-election-dnc.html}.

\bibitem{marczak2015hacking}
Bill Marczak, John Scott-Railton, and Sarah McKune.
\newblock Hacking team reloaded? {US}-based {E}thiopian journalists again
  targeted with spyware.
\newblock {\em Citizen Lab}, 9, 2015.

\bibitem{marczak2017social}
William~R Marczak and Vern Paxson.
\newblock Social engineering attacks on government opponents: Target
  perspectives.
\newblock {\em Proceedings on Privacy Enhancing Technologies},
  2017(2):172--185, 2017.

\bibitem{marczak2014governments}
William~R Marczak, John Scott-Railton, Morgan Marquis-Boire, and Vern Paxson.
\newblock When governments hack opponents: A look at actors and technology.
\newblock In {\em 23rd {USENIX} Security Symposium ({USENIX} Security 14)},
  pages 511--525, 2014.

\bibitem{matthews2017stories}
Tara Matthews, Kathleen O'Leary, Anna Turner, Manya Sleeper, Jill~Palzkill
  Woelfer, Martin Shelton, Cori Manthorne, Elizabeth~F Churchill, and Sunny
  Consolvo.
\newblock Stories from survivors: Privacy \& security practices when coping
  with intimate partner abuse.
\newblock In {\em Proceedings of the 2017 CHI Conference on Human Factors in
  Computing Systems}, pages 2189--2201, 2017.

\bibitem{mcgregor2015investigating}
Susan~E McGregor, Polina Charters, Tobin Holliday, and Franziska Roesner.
\newblock Investigating the computer security practices and needs of
  journalists.
\newblock In {\em 24th {USENIX} Security Symposium ({USENIX} Security 15)},
  pages 399--414, 2015.

\bibitem{mcgregor2016individual}
Susan~E McGregor, Franziska Roesner, and Kelly Caine.
\newblock Individual versus organizational computer security and privacy
  concerns in journalism.
\newblock {\em Proceedings on Privacy Enhancing Technologies},
  2016(4):418--435, 2016.

\bibitem{revilla2017ideal}
Melanie Revilla and Carlos Ochoa.
\newblock Ideal and maximum length for a web survey.
\newblock {\em International Journal of Market Research}, 59(5):557--565, 2017.

\bibitem{sanchez2010analysis}
Juan S{\'a}nchez-Fern{\'a}ndez, Francisco Mu{\~n}oz-Leiva, Francisco~J
  Montoro-R{\'\i}os, and Jos{\'e}~{\'A}ngel Ib{\'a}{\~n}ez-Zapata.
\newblock An analysis of the effect of pre-incentives and post-incentives based
  on draws on response to web surveys.
\newblock {\em Quality \& Quantity}, 44(2):357--373, 2010.

\bibitem{scott2016security}
John Scott-Railton.
\newblock Security for the high-risk user: separate and unequal.
\newblock {\em IEEE Security \& Privacy}, 14(2):79--87, 2016.

\bibitem{scott2017bitter}
John Scott-Railton, Bill Marczak, Claudio Guarnieri, and Masashi
  Crete-Nishihata.
\newblock Bitter sweet: Supporters of {M}exico’s soda tax targeted with {NSO}
  exploit links.
\newblock Technical report, Citizen Lab, Munk Centre for International Studies
  and University of Toronto, 2017.

\bibitem{sierra2013digital}
Jorge~Luis Sierra.
\newblock Digital and mobile security for {M}exican journalists and bloggers.
\newblock Technical report, Freedom House and the International Center for
  Journalists, 2013.
\newblock Retrieved June 25, 2020 from
  \url{https://freedomhouse.org/report/special-report/2013/digital-and-mobile-security-mexican-journalists-and-bloggers}.

\bibitem{yarosh2013shifting}
Svetlana Yarosh.
\newblock Shifting dynamics or breaking sacred traditions? {T}he role of
  technology in twelve-step fellowships.
\newblock In {\em Proceedings of the SIGCHI Conference on Human Factors in
  Computing Systems}, pages 3413--3422, 2013.

\end{thebibliography}

\end{document}